\documentclass{appolb}

\usepackage{graphicx} 
\usepackage{amssymb} 
\usepackage{amstext}
\usepackage{amsfonts} 
\usepackage{slashed} 
\usepackage[dvipsnames]{xcolor}

\newcommand{\be}{\begin{equation}} 
\newcommand{\ee}{\end{equation}}

\begin{document}

\title{Hypernuclear constraints on $\Lambda N$ and $\Lambda NN$ 
interactions\footnote{Contribution to the Dave Roper Special Volume of 
Acta Physica Polonica B.}} 
\author{Eliahu Friedman and Avraham Gal \address{Racah Institute of Physics, 
The Hebrew University \\ Jerusalem 9190400, Israel}} 

\maketitle 

\begin{abstract} 
Recent work on using density dependent $\Lambda$-nuclear optical potentials 
in calculations of $\Lambda$-hypernuclear binding energies is reviewed. 
It is found that all known $\Lambda$ binding energies in the mass range 
$16 \leq A \leq 208$ are well fitted in terms of two interaction parameters: 
one, attractive, for the spin-averaged $\Lambda N$ interaction and another 
one, repulsive, for the $\Lambda NN$ interaction. The $\Lambda N$ interaction 
term by itself overbinds $\Lambda$ hypernuclei, in quantitative agreement with 
recent findings obtained in EFT and Femtoscopy studies. The strength of the 
$\Lambda NN$ interaction term is compatible with values required to resolve 
the hyperon puzzle. 
\end{abstract} 

\section{Introduction and background}
\label{sec:intro} 

Binding energies of $\Lambda$ hyperons in single-particle states of 
$\Lambda$ hypernuclei along the periodic table have been studied since 
the 1970s~\cite{GHM16}. Although fitted by several Skyrme-Hartree-Fock 
versions, e.g., Refs.~\cite{MDG88,SH14}, a systematic study using microscopic 
density-dependent (DD) optical potentials consisting of two-body $\Lambda N$ 
and three-body $\Lambda NN$ interaction terms was lacking. The need to 
consider both terms in constructing a DD $\Lambda$-nucleus optical potential 
$V_{\Lambda}^{\rm opt}(\rho)$ was prompted by recent observations of neutron 
stars (NS) with mass exceeding twice solar mass. Such observations appeared 
to be in conflict with the expectation that $\Lambda$ hyperons in NS cores 
interacting exclusively with two-body attractive $\Lambda N$ interactions 
would soften the NS equation of state, thereby limiting NS masses to below 
$\sim$1.5 solar mass. This issue, named the `hyperon puzzle', has been 
dicussed extensively in recent years~\cite{TF20}. However, a mechanism 
of inhibiting the appearance of $\Lambda$ hyperons through a repulsive 
$\Lambda NN$ interaction apparently explains the astrophysical 
observations~\cite{GKW20}. This led us to revisit the construction 
of a proper DD $V_{\Lambda}^{\rm opt}$. 

\begin{figure}[!ht] 
\begin{center} 
\includegraphics[width=0.9\textwidth]{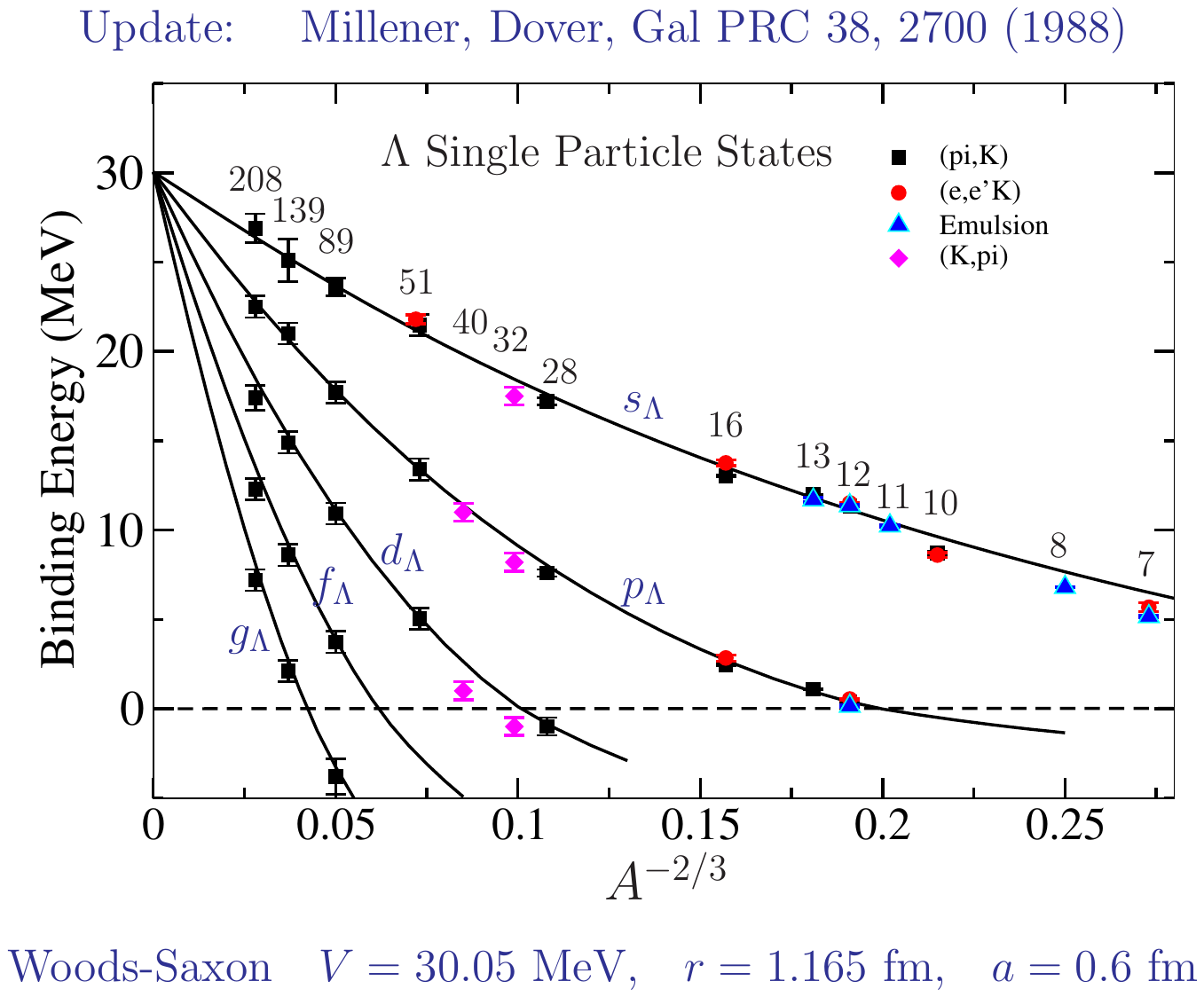} 
\caption{A three-parameter Woods-Saxon potential fit of all known $\Lambda$ 
single-particle binding energies from various experiments across the periodic 
table marked by different colors. Figure adapted from Ref.~\cite{GHM16}, 
updating the original figure in Ref.~\cite{MDG88}.} 
\label{fig:MDG88} 
\end{center} 
\end{figure} 

The $\Lambda$ binding-energy input to the present work is shown in
Fig.~\ref{fig:MDG88} which presents a three-parameter Woods-Saxon (WS) fit to
all $\Lambda$ s.p. binding energies from various experiments marked in the
figure. A limiting value of $B_{\Lambda}(A)\to 30$~MeV as $A\to\infty$ is
implied by this fit, updating the original 28~MeV value from the 1988 first
comprehensive data analysis~\cite{MDG88}. However, since the chosen WS
potential is not related directly to the nuclear density $\rho(r)$, the
remarkable fit shown in Fig.~\ref{fig:MDG88} does not tell how this 30~MeV
$\Lambda$-nuclear potential depth is split between $\Lambda N$ two-body and
$\Lambda NN$ three-body contributions.

The present work offers a brief summary and extension of recently 
published work on this topic~\cite{FG23a,FG23b}. In Sect.~\ref{sec:method}, 
$V_{\Lambda}^{\rm opt}$ is constructed in terms of nuclear densities based 
on nuclear sizes. Sect.~\ref{sec:Lfits} shows optical-potential predictions 
obtained, first by fitting its $\Lambda N$ and $\Lambda NN$ strength 
parameters to the $1s_\Lambda$ and $1p_\Lambda$ binding energies in 
$^{16}_{~\Lambda}$N, and then by a $\chi^2$ fit to $1s_\Lambda$ and 
$1p_\Lambda$ single-particle (s.p.) binding energies shown in 
Fig.~\ref{fig:MDG88} across the periodic table. A need to suppress the 
$\Lambda NN$ interaction term when one of the two nucleons is an `excess' 
neutron is observed, apparently related to its isospin dependence. 
Sect.~\ref{sec:iso} demonstrates a way to substantiate the isospin dependence 
of the $\Lambda NN$ term in $^{40,48}$Ca($e,e'K^+)^{40,48}_{~~~~\Lambda}$K 
electroproduction experiments scheduled at JLab. The concluding 
Sect.~\ref{sec:sum} presents a brief discussion and summary of our optical 
potential methodology in comparison with other approaches. In particular, 
the strength of the repulsive $\Lambda NN$ interaction term of 
$V_{\Lambda}^{\rm opt}$ found here is compatible with that required to resolve 
the hyperon puzzle~\cite{GKW20}.

\section{DD optical potentials} 
\label{sec:method} 

The optical potential $V_{\Lambda}^{\rm opt}(\rho)=V_{\Lambda}^{(2)}(\rho)+
V_{\Lambda}^{(3)}(\rho)$ consists of two-body $\Lambda N$ and three-body 
$\Lambda NN$ interaction terms:
\begin{equation}
V_{\Lambda}^{(2)}(\rho) = -\frac{4\pi}{2\mu_{\Lambda}}\frac{f^{(2)}_A\,b_0}
{1+\frac{3k_F}{2\pi}f^{(2)}_A\,b_0}\,\rho 
\label{eq:V2}
\end{equation}
\begin{equation}
V_{\Lambda}^{(3)}(\rho) = +\frac{4\pi}{2\mu_{\Lambda}}f^{(3)}_A\,B_0\,
\frac{\rho^2}{\rho_0}, 
\label{eq:V3}
\end{equation}
with $b_0$ and $B_0$ strength parameters in units of fm ($\hbar=c=1$).
In these expressions, $A$ is the mass number of the {\it nuclear core}
of the hypernucleus, $\rho$
is a nuclear density normalized to $A$, $\rho_0=0.17$~fm$^{-3}$ stands for
nuclear-matter density, $\mu_{\Lambda}$ is the $\Lambda$-nucleus reduced 
mass and $f^{(2,3)}_A$ are kinematical factors involved in going from the 
$\Lambda N$ and $\Lambda NN$ c.m. systems, respectively, to the 
$\Lambda$-nucleus c.m. system:
\begin{equation}
f^{(2)}_A=1+\frac{A-1}{A}\frac{\mu_{\Lambda}}{m_N},\,\,\,\,\,\,
f^{(3)}_A=1+\frac{A-2}{A}\frac{\mu_{\Lambda}}{2m_N}.
\label{eq:fA}
\end{equation}
The novelty of this form of $V_{\Lambda}^{\rm opt}(\rho)$ is that its $V^{(2)}$ 
component accounts explicitly for Pauli correlations through its dependence on 
the Fermi momentum $k_F=(3{\pi^2}\rho/2)^{1/3}$, which affects strongly the 
balance between the derived values of potential parameters $b_0$ and $B_0$. 
In contrast, introducing Pauli correlations also in $V_{\Lambda}^{(3)}$ 
is found to make little difference, which is why it is skipped in 
Eq.~(\ref{eq:V3}). This form for including Pauli correlations was suggested 
in Ref.~\cite{WRW97} and practised by us since 2013 in $K^-$ atom~\cite{FG13} 
and $\eta$-nuclear studies~\cite{FGM13} within a Jerusalem-Prague 
collaboration as reviewed in Refs.~\cite{FG18a,FG18b}. We note that the 
$\Lambda NN$ potential term $V_{\Lambda}^{(3)}(\rho)$ derives mostly 
from OPE diagrams with $\Sigma NN$ and $\Sigma^\ast NN$ intermediate 
states~\cite{GSD71}, whereas $NN$ short-range correlations are estimated to 
affect the derived value of $B_0$ by a few percent at most. Finally, recall 
that the low-density limit of $V_{\Lambda}^{\rm opt}$ requires $b_0$ to be 
identified with the c.m. $\Lambda N$ spin-averaged scattering length, taken 
positive here. 

Regarding the nuclear densities $\rho(r) = \rho_p(r) + \rho_n(r)$ used in 
$V_{\Lambda}^{\rm opt}$, it is essential to ensure that the radial extent of 
the densities, e.g., their r.m.s. radii, follow closely values derived
from experiment. Here we relate the proton densities to the corresponding 
charge densities where proton-charge finite size and recoil effects are 
included. This is equivalent to assigning some finite range to the 
$\Lambda N$ interaction. For the lightest elements in our database we used 
harmonic-oscillator type densities, assuming the same radial parameters
also for the corresponding neutron densities \cite{Elton61}. For species
beyond the nuclear $1p$ shell we used two-parameter and three-parameter
Fermi distributions normalized to $Z$ for protons and $N=A-Z$ for neutrons,
derived from nuclear charge distributions assembled in Ref.~\cite{AM13}.
For medium-weight and heavy nuclei, the r.m.s. radii of neutron density
distributions assume larger values than those for protons. Furthermore, 
once neutrons occupy single-nucleon orbits beyond those occupied by protons, 
it is useful to represent the nuclear density $\rho(r)$ as
\begin{equation}
\rho(r)=\rho_{\rm core}(r)+\rho_{\rm excess}(r),
\label{eq:exc1}
\end{equation}
where $\rho_{\rm core}$ refers to the $Z$ protons plus the charge-symmetric
$Z$ neutrons occupying the same nuclear `core' orbits, and $\rho_{\rm excess}$
refers to the $(N-Z)$ `excess' neutrons associated with the nuclear periphery.

\section{Optical-potential fits to $\Lambda$ hypernuclear binding energies} 
\label{sec:Lfits} 

Motivated by the simple $1p$ proton-hole structure of the $^{15}$N nuclear 
core of $^{16}_{~\Lambda}$N, which removes most of the uncertainty from 
spin-dependent $\Lambda N$ and $\Lambda NN$ interactions, we started our 
optical-potential study of $\Lambda$ hypernuclei by fitting $B_{\Lambda}(1s)$ 
and $B_{\Lambda}(1p)$ in $^{16}_{~\Lambda}$N. The fit parameters are 
$b_0$=1.53~fm, $B_0$=0.22~fm, and the corresponding partial potential 
depths and total depth at nuclear-matter density $\rho_0=0.17$~fm$^{-3}$ 
for $A\to\infty$ are  
\begin{equation} 
D^{(2)}_\Lambda=-39.3~{\rm MeV},\,\,\,D^{(3)}_\Lambda=+13.1~{\rm MeV},\,\,\, 
D_\Lambda=-26.2~{\rm MeV}.  
\label{eq:total0}
\end{equation} 

\begin{figure}[!ht] 
\begin{center} 
\includegraphics[width=0.8\textwidth]{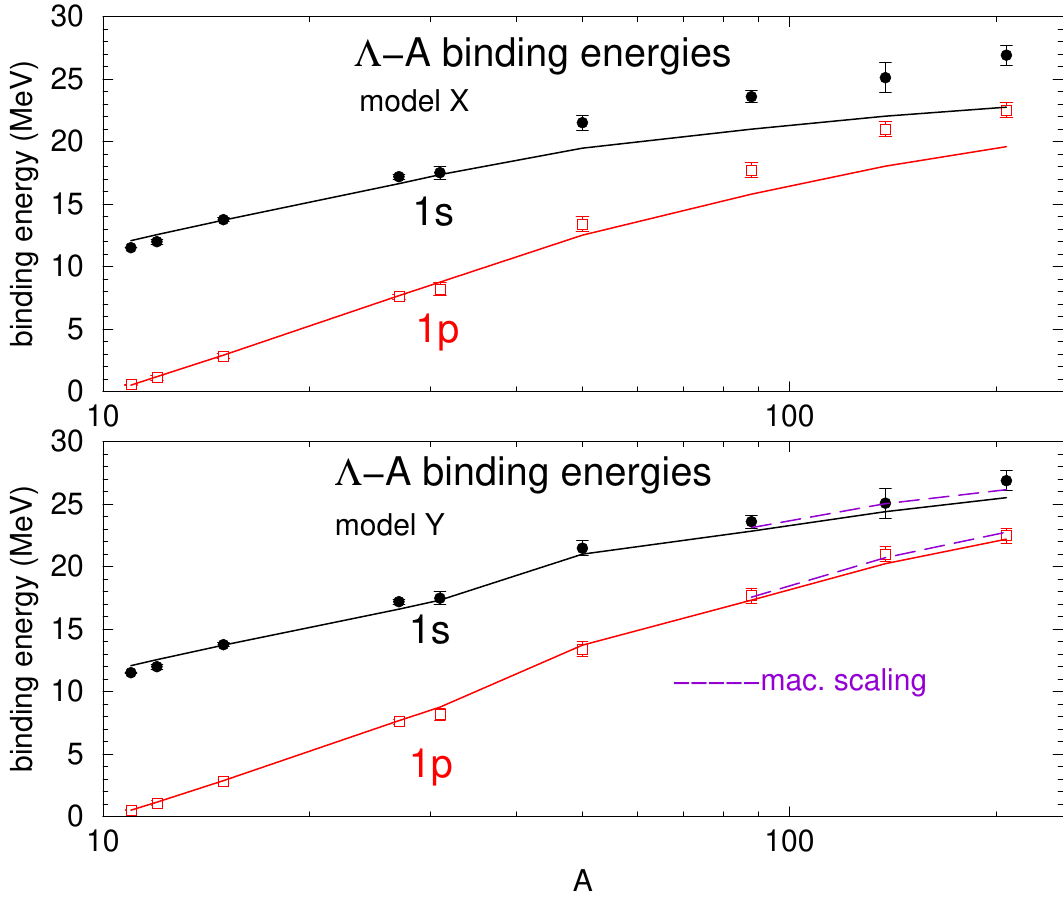} 
\caption{$B_{\Lambda}^{1s,1p}(A)$ values across the periodic table, 
$12\leq A \leq 208$ as calculated in models X (upper) and Y (lower), 
compared with data points, including uncertainties. $^{16}_{~\Lambda}$N 
is the third point on each line. Continuous lines connect calculated values. 
Figure updating Fig.~3 in Ref.~\cite{FG23a}. The upper part, model X, uses 
the full $\rho^2$ term. The lower part, model Y, replaces $\rho^2$ with 
a reduced form, decoupling $(N-Z)$ excess neutrons from $2Z$ symmetric-core 
nucleons, see text. The dashed lines are for $\rho^2$ replaced with $F\rho^2$, 
with a suppression factor $F$, Eq.~(\ref{eq:F}).} 
\label{fig:XY} 
\end{center} 
\end{figure} 

Next, we show in the top part of Fig.~\ref{fig:XY} (model X) results of using 
the $\Lambda$ optical-potential strength parameters $b_0$ and $B_0$ of 
Eqs.~(\ref{eq:V2}) and (\ref{eq:V3}), determined by fitting to 
$^{16}_{~\Lambda}$N $1s_\Lambda$ and $1p_\Lambda$ binding energies, 
in calculations of {\it all} $1s_\Lambda$ and $1p_\Lambda$ pairs of 
binding energies along the periodic table. Clearly seen is underbinding of 
$1s_\Lambda$ and $1p_\Lambda$ states in the heavier hypernuclei, which could 
result from treating equally all $NN$ pairs, including pairs where one nucleon 
is in the nuclear `core' while the other is an `excess' neutron. Removing this 
bilinear term from $\rho ^2$, using Eq.~(\ref{eq:exc1}), we replace $\rho ^2$ 
beyond $^{40}$Ca with 
\begin{equation}
\rho_{\rm core}^2+\rho_{\rm excess}^2\rightarrow(2\rho_p)^2+(\rho_n-\rho_p)^2,
\label{eq:exc2}
\end{equation}
approximated in terms of the available densities $\rho_p$ and $\rho_n$. This 
prescription is suggested by the ${\vec\tau}_1\cdot{\vec\tau}_2$ isospin 
dependence that arises from intermediate $\Sigma$ and $\Sigma^\ast$ hyperons 
in the $\Lambda NN$ OPE interaction~\cite{GSD71}. The associated effect of 
weakening the repulsive $\Lambda NN$ term beyond $^{40}$Ca is seen in the 
lower part of Fig.~\ref{fig:XY} (model Y).

In order to simplify things, one may multiply $\rho^2$ by a suppression factor 
\begin{equation}
F=\frac{(2Z)^2+(N-Z)^2}{A^2}.
\label{eq:F}
\end{equation}
This suppression factor, approximately the ratio of the volume integral of 
$(2\rho_p)^2+(\rho_n-\rho_p)^2$ to that of $\rho^2$, becomes significant for 
heavy hypernuclei, with as small value as $F$=0.67 for Pb. Results using 
$F\rho^2$ instead of Eq.~(\ref{eq:exc2}) are shown in the lower part of the 
figure as dashed lines (`mac. scaling'), leading to almost identical values 
for these two options. 

\begin{figure}[!ht] 
\begin{center} 
\includegraphics[width=0.7\textwidth]{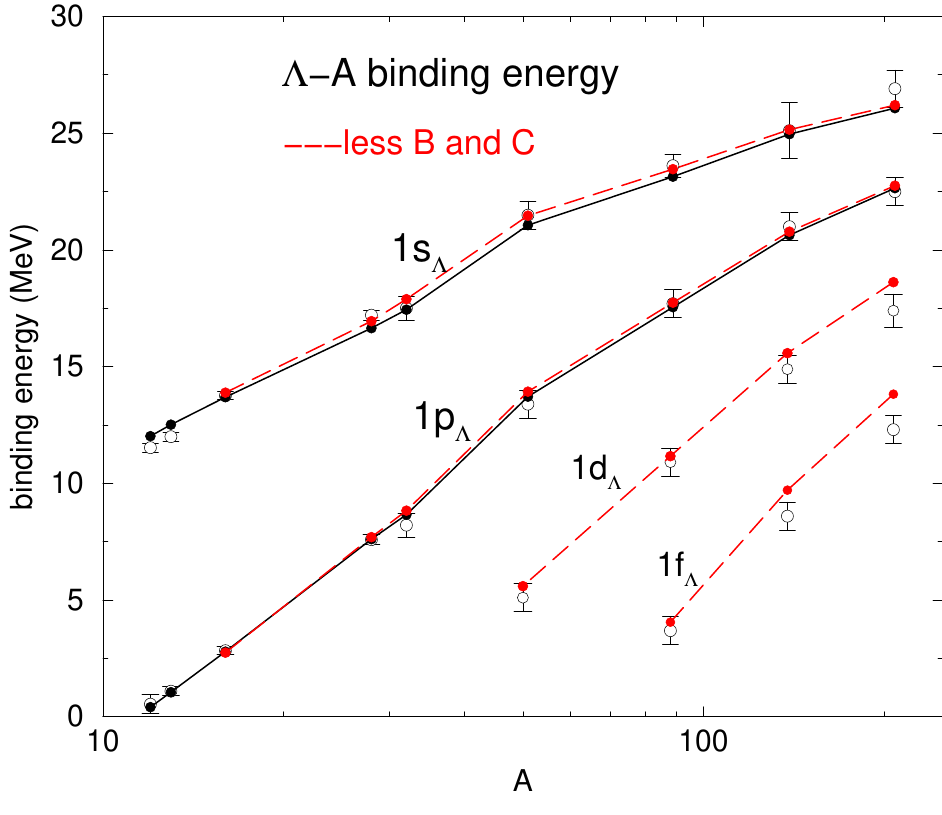} 
\caption{$\chi^2$ fits to the full 1$s_\Lambda$ and $1p_\Lambda$ 
data (solid black lines) and when excluding $^{12}_{~\Lambda}$B and 
$^{13}_{~\Lambda}$C (dashed red lines). Also shown are {\it predictions} 
of 1$d_\Lambda$ and 1$f_\Lambda$ binding energies for the latter choice.} 
\label{fig:LambdaALL} 
\end{center} 
\end{figure} 

At the next stage we performed full $\chi^2$ fits with the suppression factor 
$F$ included for the four heaviest species. Figure \ref{fig:LambdaALL} shows 
several fits to the ($1s_{\Lambda}$, $1p_{\Lambda}$) $B_{\Lambda}$ data 
where black solid lines show fits to the full data set and open circles 
with error bars mark $B_{\Lambda}$ data points listed in Table IV of 
Ref.~\cite{GHM16}, including experimental uncertainties, and summarized 
in Table 2 of Ref.~\cite{FG23b}. It is clearly seen that the $1s_{\Lambda}$ 
states in $^{12}_{~\Lambda}$B and $^{13}_{~\Lambda}$C do not fit into the 
otherwise good agreement with experiment for the heavier species. The red 
dashed lines show a very good fit obtained upon excluding these two light 
elements from the $B_{\Lambda}$ data set. In fact, the potential parameters 
$b_0$ and $B_0$ of Eqs.~(\ref{eq:V2},\ref{eq:V3}) are hardly affected by 
the $^{12}_{~\Lambda}$B and $^{13}_{~\Lambda}$C $B_\Lambda$ input. 
The fit parameters are $b_0=1.446\pm 0.088$~fm, $B_0=0.193\pm 0.022$~fm, 
with 100\% correlation between the two parameters. 
The corresponding fully correlated partial potential depths, and the full one, 
at nuclear-matter density $\rho_0=0.17$~fm$^{-3}$ are (in MeV):
\begin{equation}
D^{(2)}_\Lambda=-38.6\mp 0.8,\,\,\,\,\,\,
D^{(3)}_\Lambda=11.5\pm 1.4,\,\,\,\,\,\,
D_\Lambda=-27.1\pm 0.6.
\label{eq:total1}
\end{equation} 

Also included in Fig.~\ref{fig:LambdaALL} are {\it predictions} of
$1d_\Lambda$ and $1f_\Lambda$ binding energies made with these parameters.
Although it is not expected that such states will be well described by the
same potential, owing to overlooked secondary effects such as non-local terms,
it is seen that while slight overbinding of the calculated energies appears
for the heavier species, the present optical potential reproduces quite well
the four lowest $\Lambda$ single-particle states in neutron-rich hypernuclei.
It is therefore of interest to repeat the $\chi^2$ process on the whole
set of experimental binding energies of $\Lambda$ single-particle states,
a total of 21 binding-energy values between $^{16}_{~\Lambda}$N and
$^{208}_{~~\Lambda}$Pb. The resulting $\chi^2$ per degree of freedom is then
0.95 (compared to 0.6 from a fit to only 14 binding-energy values for the
$1s_\Lambda$ and $1p_\Lambda$ states) and the potential parameters are 
$b_0=1.32\pm 0.071$~fm, $B_0=0.164\pm 0.020$~fm, and are 100\% correlated. 
The corresponding fully correlated partial potential depths and the full one
at nuclear-matter density $\rho_0=0.17$~fm$^{-3}$ are (in MeV):
\begin{equation}
D^{(2)}_\Lambda=-37.4\mp 0.7,\,\,\,\,\,\,
D^{(3)}_\Lambda=9.8\pm 1.2,\,\,\,\,\,\,
D_\Lambda=-27.6\pm 0.5.
\label{eq:total2}
\end{equation}
These values are in agreement with those in Eq.~(\ref{eq:total1}) based only 
on $1s_\Lambda$ and $1p_\Lambda$ states. The uncertainties in the parameter
values quoted above are statistical only. To estimate systematic effects
within the adopted model we repeated the analysis with slightly modified
nuclear densities such as obtained when unfolding the finite size of the
proton. Values of $b_0$ came out unchanged whereas values of $B_0$ increased
typically by 0.015~fm. The total potential depth at $\rho_0=0.17$~fm$^{-3}$
changed to $D_\Lambda=-26.8\pm 0.4$~MeV, suggesting a systematic uncertainty
of somewhat less than 1 MeV for this value.

\section{Test of isospin dependence} 
\label{sec:iso} 

\begin{figure}[!h] 
\begin{center} 
\includegraphics[width=0.7\textwidth]{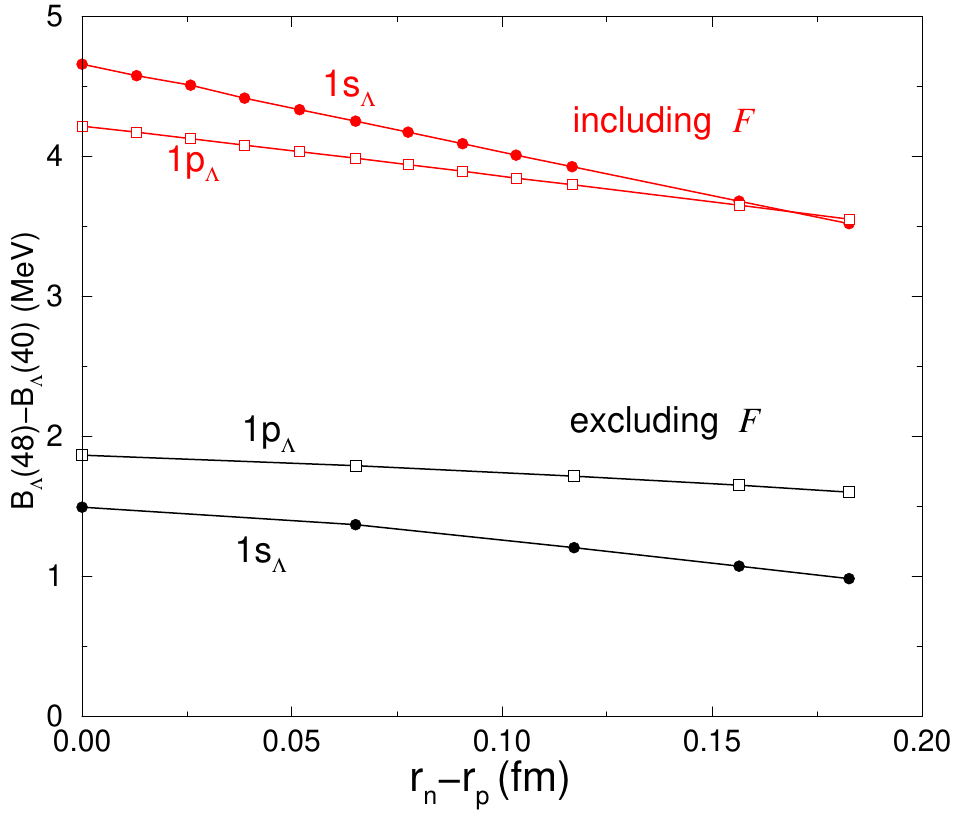} 
\caption{$B_\Lambda(^{48}_{~\Lambda}{\rm K})-B_\Lambda(^{40}_{~\Lambda}{\rm K} 
)$ values for $1s_{\Lambda}$ and $1p_{\Lambda}$ states, with and without 
applying the suppression factor $F$, as a function of the neutron-skin of 
$^{48}_{~\Lambda}$K, see text.} 
\label{fig:iso} 
\end{center} 
\end{figure} 

Having imposed on our $B_\Lambda$ optical-potential fits the isospin-related 
suppression factor $F$, Eq.~(\ref{eq:F}), we discuss here a test case which 
demonstrates its effect on $\Lambda$ hypernuclear spectra. Fig.~\ref{fig:iso} 
shows calculated differences of $B_\Lambda$ values for the $1s_{\Lambda}$ and 
$1p_{\Lambda}$ states between $^{48}_{~\Lambda}$K and $^{40}_{~\Lambda}$K 
as a function of the neutron skin $r_n-r_p$ in $^{48}_{~\Lambda}$K. 
The upper/lower part of the figure shows predictions made using 
$V_{\Lambda}^{\rm opt}$ upon including/excluding the suppression factor $F$. 
Regardless of the chosen value of $r_n-r_p$, the effect of applying $F$ is 
about 2.5~MeV for the $1s_{\Lambda}$ state and more than 2~MeV for the 
$1p_{\Lambda}$ state, both are within reach of the ($e,e'K^+$) approved 
JLab E12-15-008 experiment on $^{40,48}$Ca targets~\cite{Nakamura22}. 
For $F=1$, our calculated $B_\Lambda$ values are close to those 
calculated in Refs.~\cite{IYR17,Byd23,Byd25}.

\section{Discussion and summary} 
\label{sec:sum} 

\begin{table}[!b]
\begin{center}
\begin{tabular}{cccc}
\hline
Model & $D_{\Lambda}^{(2)}$~(MeV) & $D_{\Lambda}^{(3)}$~(MeV) &
$D_{\Lambda}$~(MeV) \\
\hline
with $PC$~\cite{FG23a} & $-$39.3 & $+$13.1 & $-$26.2 \\ 
w/o $PC$~\cite{MDG88}  & $-$57.8 & $+$31.4 & $-$26.4 \\ 
\hline 
\end{tabular} 
\caption{Comparison of two-body, three-body and total $\Lambda$-nucleus 
potential depths at nuclear matter density $\rho_0=0.17$~fm$^{-3}$ obtained 
by fitting $b_0$ and $B_0$ to $B_{\Lambda}(1s)$ and $B_{\Lambda}(1p)$ in 
$^{16}_{~\Lambda}$N with and without considering Pauli correlations; 
see text.} 
\label{tab:PC} 
\end{center} 
\end{table} 

First, we discuss briefly the effect of Pauli correlations entered in 
$V_{\Lambda}^{(2)}(\rho)$, Eq.~(\ref{eq:V2}), by the renormalization factor 
$PC=(1+\frac{3k_F}{2\pi}f^{(2)}_A\,b_0)^{-1}$. In Table~\ref{tab:PC} we list 
values of partial potential depths $D_{\Lambda}^{(2)}$ and $D_{\Lambda}^{(3)}$ 
determined by fitting $B_{\Lambda}(1s)$ and $B_{\Lambda}(1p)$ in 
$^{16}_{~\Lambda}$N with/without $PC$. The total depth, as expected, 
is practically the same in both derivations. The difference in the partial 
potential depths is traced to the treatment of Pauli correlations. Expanding 
$PC$ in powers of $k_F \propto {\rho}^{1/3}$, its repulsive $\rho$ term along 
with the overall $\rho$ factor in $V_{\Lambda}^{(2)}$ produce a repulsive 
$\rho^2$ term in $V_{\Lambda}^{(2)}(\rho)$, thereby enhancing 
$V_{\Lambda}^{(3)}(\rho)\propto \rho^2$ which then boosts $D_{\Lambda}^{(3)}$. 
Finally, to keep the total depth $D_{\Lambda}$ about the same, 
$|D_{\Lambda}^{(2)}|$ must increase too. A similar effect of boosting both 
values of $|D_{\Lambda}^{(2)}|$ and $D_{\Lambda}^{(3)}$ is seen also in 
Skyrme-Hartree-Fock (SHF) analyses of $\Lambda$ s.p. binding energies, 
e.g. Refs.~\cite{MDG88,SH14}, which treat Pauli correlations incompletely. 

\begin{table}[!ht]  
\begin{center} 
\begin{tabular}{c|c|c|c} 
\hline
Model & $D_{\Lambda}^{(2)}$ & $D_{\Lambda}^{(3)}$ & $D_{\Lambda}$ \\ 
\hline 
Nijmegen ESC16,16$^+$~\cite{ESC19} & $-$43.7 & $+$5.8 & $-$37.9 \\
EFT NLO19~\cite{HMN20} & $-$39 to $-$29 & -- & -- \\
NLO19 + $\bf{10}$ dominance~\cite{GKW20} & $\approx -36$ & $\approx +10$ & \\
EFT N$^2$LO~\cite{HMNL23} & $-$33 to $-$38 & -- & -- \\
Femtoscopy~\cite{MHMS24} & $-36.3{\pm 1.3} {^{+2.5}_{-6.2}}$
& -- & -- \\ 
\hline 
$V_{\Lambda}^{\rm opt}$ [present] (${\cal N}$=14) & $-38.6 \mp 0.8$ & 
$11.5 \pm 1.4$ & $-27.1 \pm 0.6$ \\  
$V_{\Lambda}^{\rm opt}$ [present] (${\cal N}$=21) & $-37.4 \mp 0.7$ & 
$9.8 \pm 1.2$ & $-27.6 \pm 0.5$ \\
\hline
\end{tabular}
\caption{Two-body, three-body and total $\Lambda$-nucleus potential depths
(in MeV) at nuclear matter density $\rho_0=0.17$~fm$^{-3}$ from several model 
calculations and from hypernuclear binding-energy data 
(${\cal N}$ stands for the number of data points).}
\label{tab:results} 
\end{center} 
\end{table} 

Results of various $\Lambda$-nuclear potential-depth calculations are listed 
in Table~\ref{tab:results}, divided into two groups. The first group consists 
of recent calculations, mostly EFT, confirming that $\Lambda N$ interactions 
overbind $\Lambda$ hypernuclei by as much as about 10~MeV, a value consistent 
with a $\Lambda NN$ repulsive contribution of order 10~MeV in order to satisfy 
the total potential depth value $D_{\Lambda}\approx -27$~MeV suggested 
by the overall $B_\Lambda$ data. The second group demonstrates our own 
$V_{\Lambda}^{\rm opt}$ determination of $\Lambda$-nuclear partial potential 
depths, Eqs.~(\ref{eq:total1}) and (\ref{eq:total2}). Our fitted value of 
the $\Lambda NN$ potential depth, $D_{\Lambda}^{(3)}\approx 10$~MeV, agrees 
with the $\Lambda NN$ potential strength derived in the Gerstung-Kaiser-Weise 
calculation~\cite{GKW20} which appears to resolve the hyperon puzzle.


\end{document}